\documentclass{article}
\pdfoutput=1
\usepackage[affil-it]{authblk}
\usepackage{url}
\usepackage{amsmath}
\usepackage{amsfonts}
\usepackage{graphicx}
\usepackage{algorithm}
\usepackage{algorithmic}
\usepackage{textcomp}
\usepackage{booktabs}
\usepackage{array}
\usepackage{caption}
\usepackage{subcaption}
\usepackage{pifont}
\usepackage[UKenglish]{babel}
\DeclareMathOperator*{\argmin}{\arg\!\min}

\begin{document}
	\title{Can Planning Images Reduce Scatter in Follow-Up Cone-Beam CT?}
	\author{Jonathan Mason\textsuperscript{1}%
		\thanks{Electronic address: \texttt{j.mason@ed.ac.uk}; Corresponding author} }
	\author{Alessandro Perelli\textsuperscript{1}}\author{William Nailon\textsuperscript{2}}\author{\\Mike Davies\textsuperscript{1}}
\affil{\textsuperscript{1}Institute for Digital Communications, University of Edinburgh,\\ EH9 3JL, UK}
\affil{\textsuperscript{2}Oncology Physics Department, Edinburgh Cancer Centre, \\Western General Hospital, Edinburgh, EH4 2XU, UK}

\maketitle

\begin{abstract}
	Due to its wide field of view, cone-beam computed tomography (CBCT) is plagued by large amounts of scatter, where attenuated photons hit the detector, and corrupt the linear models used for reconstruction. Given that one can generate a good estimate of scatter however, then image accuracy can be retained. In the context of adaptive radiotherapy, one usually has a low-scatter planning CT image of the same patient at an earlier time. Correcting for scatter in the subsequent CBCT scan can either be self consistent with the new measurements or exploit the prior image, and there are several recent methods that report high accuracy with the latter. In this study, we will look at the accuracy of various scatter estimation methods, how they can be effectively incorporated into a statistical reconstruction algorithm, along with introducing a method for matching off-line Monte-Carlo (MC) prior estimates to the new measurements. Conclusions we draw from testing on a neck cancer patient are: statistical reconstruction that incorporates the scatter estimate significantly outperforms analytic and iterative methods with pre-correction; and although the most accurate scatter estimates can be made from the MC on planning image, they only offer a slight advantage over the measurement based scatter kernel superposition (SKS) in reconstruction error.
\end{abstract}

\section{Introduction}
Cone-beam computed tomography (CBCT) is an imaging modality that is seeing increased use for image guidance procedures, such as radiation therapy \cite{Button2010}. A key challenge of this geometry is the vast quantities of scattered photons that reach the detector \cite{Siewerdsen2001}, and contaminate other line-of-sight measurements. Usually in this context however, one has a planning scan of the same patient from a more accurate CT acquisition, such as a helical fan-beam, which has significantly lower scatter due to better collimation and narrower field-of-view.  Typical approaches to CBCT scatter correction either form a self-consistent model based solely on the new measurements \cite{Poludniowski2009a}\cite{Love1987}\cite{Boellaard1997}\cite{Sun2010}, or exploit the prior image as a basis for its estimation \cite{Nui2010}\cite{Marchant2008}\cite{Xu2015}.

In this study, we will adopt the perspective that various scatter models allow one to make an estimate of its expectation, given a set of new CBCT projections, and potential access to a low-scatter prior CT image. We will then investigate two key aspects: how accurate can scatter be estimated in a moderate and low dose settings, and what is the robustness of various methods in this challenging scenario; and how should these estimates be incorporated into reconstruction.

In the first case, the scatter estimation methods we will look at fall into two distinct classes---methods that are blind to the planning image, and methods that exploit it. In our `prior blind' category are a view dependent uniform estimation \cite{Boellaard1997}, the scatter kernel superposition (SKS) \cite{Love1987} and fast asymmetric SKS (fASKS) \cite{Sun2010}, along with simulating the scatter through a Monte-Carlo (MC) engine on a preliminary reconstruction with the fast Feldkamp--Davis--Kress (FDK) \cite{Feldkamp1984} algorithm. Conversely, based upon a rigid registration of the planning image, we will look at the effectiveness of taking scatter as a smooth projection difference \cite{Nui2010}\cite{Park2015}, and from using the MC engine on this registered plan. Here, we will look at both calculating the planning MC estimate on-line, after registration \cite{Xu2015}, along with the notion of matching an off-line pre-calculated estimate to the measurements.

In the second case, in the subsequent reconstruction with each of the estimation methods, we make a distinction between `scatter correction', where measurements a pre-processed to remove its effect, and `scatter-aware inference', where the imaging operates based on the raw uncorrected measurements and knowledge of the scatter estimate. Most popular `analytic' and `iterative' techniques, such as FDK and PWLS \cite{Fessler2014}, fall into the former category. Performing the inference, although more challenging due to its non-linearity and nonconvexity, represents a more accurate data model that may mitigate reconstruction artefacts and errors.

We begin this article with relevant background material, where we explain the system model in Section~\ref{sec:model}, an overview of scatter estimation methods in Section~\ref{sec:scat}, and standard reconstruction based on pre-corrected measurements in Section~\ref{sec:recon}. Next, we give details of matching an off-line planning MC scatter given a rigid translation of the specimen in Section~\ref{sec:match}, along with the model for statistical inference reconstruction in Section~\ref{sec:scat_est}.

From a dataset derived from repeat CT images of a neck caner patient, we then evaluate both the scatter estimation accuracy and reconstruction error with the range the methods under test. The results are then presented in Section~\ref{sec:results}. 

\section{Background}
%
%

\subsection{System Description} \label{sec:model}
The system we will directly study in this work is a circular scan CBCT. This consists of a point source and flat panel detector, which rotate throughout $360^\circ$ around the specimen where photons are emitted and measured after a fixed angular increment. Assuming that recorded x-rays are drawn from independent Poisson distributions given as \cite{Elbakri2002}\cite{Chang2014}, then we can write the distribution using monoenergetic Beer-Lambert law with additive scatter component as
\begin{equation} \label{equ:stat_model}
y_k \sim \operatorname{Poisson}\left\{b_k\exp(-\left[\boldsymbol{\Phi}\boldsymbol{\mu}\right]_i)+s_k \right\} \mbox{ for }  k=1,\dots,N,
\end{equation}
where $\hat{\boldsymbol{y}}\in\mathbb{R}^N$ is a column vector of measurements, $\boldsymbol{b}\in\mathbb{R}^N$ is a vector of input source fluxes, $\boldsymbol{\mu}\in \mathbb{R}^M$ is a vector of linear attenuation coefficients, and $\boldsymbol{\Phi}\in\mathbb{R}^{N\times M}$ is a system matrix describing the path of each ray through the specimen and onto the detector.

\subsection{Estimating Scatter} \label{sec:scat}
In this section, we give brief overviews of the various scatter estimation techniques that we evaluate in this study.

\subsubsection{Uniform:}
A simple method that calculates a constant scatter at each projection angle, we denote as `uniform' \cite{Boellaard1997}\cite{Rit2014}. Here, using the assumption that a scatter-to-primary ratio (SPR) is known a-priori, along with a distinction between air and object containing projections, we can write
\begin{equation} \label{equ:uniform}
\boldsymbol{s}\{i\} = \frac{\mathrm{SPR}}{N_\mathrm{air}(i)}\sum_{k\in\mathcal{C}_\mathrm{air}(i)} y(k) \mbox{ for }  i=1,\dots,P,
\end{equation}
where $\mathcal{C}_\mathrm{air}(i)$ is the set of air containing measurements at the $i^\mathrm{th}$ angle, where the set satisfies $y(k)\geq t_\mathrm{air}\, \forall k\in\mathcal{C}_\mathrm{air}$ with some scalar threshold $t_\mathrm{air}$. This essentially calculates the mean scatter given a constant ratio. In practice, this SPR can be found by observing the magnitude of signal in the air region with and without a specimen present, and assume the difference is scatter. To ensure the scatter is less than the minimum value in a given projection, a non-negativity constraint can be added \cite{Rit2014}.

\subsubsection{SKS/ASKS:}
The scatter kernel superposition (SKS) \cite{Love1987} and asymmetric SKS (ASKS) \cite{Sun2010}, perform estimation as a convolution of the scatter free incident beam with an appropriate kernel. Since the incident beam is itself unknown, the methods iteratively estimate this as the difference between raw measurements and updated scatter estimate from the previous iteration. Due to the ability calculate convolution rapidly through the FFT, both of these methods are relatively fast, especially when the projections are sub-sampled. Although they implicitly model the scatter media as homogeneous, the estimates are accurate in practice.

\subsubsection{Diff. filt.:}
A simple concept for predicting the scatter contribution based upon a planning image is though the smooth difference---`diff. filt.'---between CBCT measurements and projections of a registered plan \cite{Nui2010}. This model can be expressed as
\begin{equation} \label{equ:diff_filt}
\boldsymbol{s} = \boldsymbol{\mathcal{F}}(\boldsymbol{y}-\boldsymbol{b}\odot\boldsymbol{\exp}(-\boldsymbol{\Phi}\boldsymbol{\mu}_\mathrm{reg})),
\end{equation}
where $\boldsymbol{\mathcal{F}}$ is a projection-wise filter, or set of filters---subsequent median and Gaussian filters are used in \cite{Nui2010}---and $\boldsymbol{\mu}_\mathrm{reg}$ is the registered planning CT onto a preliminary reconstruction of the CBCT measurements. 

\subsubsection{Monte-Carlo (MC):}
Monte-Carlo scatter estimation techniques essentially draw a number of samples from an accurate probabilistic model of physical interactions. Given that the model is a faithful representation of reality, then the true expectation of scatter can be found with infinite samples.

We denote this process as
\[
\tilde{\boldsymbol{s}} \sim \mathcal{MC}(\boldsymbol{b},\boldsymbol{\Phi},\boldsymbol{\mu}_\mathrm{est.},N_\mathcal{MC}),
\]
where $\tilde{\boldsymbol{s}}$ is the estimate after $N_\mathcal{MC}$ photon simulations distributed throughout several projection angles and $\boldsymbol{\mu}_\mathrm{est.}$ is the image onto which the estimation is based. We will test the ability of estimating MC scatter onto both a preliminary FDK with appropriate prior blind estimation, such as SKS, and onto the planning image.

If the MC simulation is made after the measurements are taken, then it may be appropriate to sub-sample both the image $\tilde{\boldsymbol{s}}$ and the number of photons, in order to complete the calculation on-line \cite{Xu2015}. In the off-line setting, there is no immediate limitation on computational time, as it can be performed days or weeks ahead of the follow-up CBCT. Eventual matching of this off-line estimate is detailed in Section~\ref{sec:match}.

\subsection{Reconstruction from Scatter Correction} \label{sec:recon}
In most cases, reconstruction is performed by inferring the attenuation coefficient given the model in (\ref{equ:stat_model}), which follows an effective `correction' of scatter. In the crudest form, this involves simply subtracting the scatter estimate from the measurements. An advantage of pre-correcting for scatter in this manner, allows a linear system to be exposed and solved, of the form
\begin{equation} \label{equ:lin-model}
\boldsymbol{p} = \boldsymbol{\Phi}\boldsymbol{\mu}+\boldsymbol{n},
\end{equation}
where $\boldsymbol{n}\in\mathbb{R}^{N}$ is noise, $\boldsymbol{p}\in\mathbb{R}^{N}$ is the linearised projection, calculated by
\begin{equation} \label{equ:lin-approx}
p_i = \log\left(\frac{b_i}{y_i-s_i}\right) \mbox{ for }  i=1,...,N,
\end{equation}
where it can be solved by analytic filtered back-projection methods, such as FDK for CBCT \cite{Feldkamp1984}, or with iterative methods, that approximate the noise model in (\ref{equ:stat_model}) and incorporate regularisation, such as penalised weighted-least-squares (PWLS). Reconstruction through PWLS involves solving the problem
\begin{equation} \label{equ:pwls}
\hat{\boldsymbol{\mu}} = \argmin_{\boldsymbol{\mu}}  \left(\boldsymbol{\Phi}\boldsymbol{\mu}-\boldsymbol{p}\right)^T\boldsymbol{W}\left(\boldsymbol{\Phi}\boldsymbol{\mu}-\boldsymbol{p}\right) + \lambda R(\boldsymbol{\mu}),
\end{equation}
where $\boldsymbol{W}\in\mathbb{R}^{N\times N}$ is a diagonal weighting matrix with entries $w_{ii} = (y_i-s_i)^2/y_i$, $R(\boldsymbol{\mu})$ is a regularisation function to promote desirable structure in $\boldsymbol{\mu}$, and $\lambda$ is usually a scalar constant trade-off between data fit and regularisation.

\section{Method} \label{sec:method}

\subsection{Off-line Scatter Matching} \label{sec:match}
\begin{figure}
	\centering
	\includegraphics[width=\linewidth]{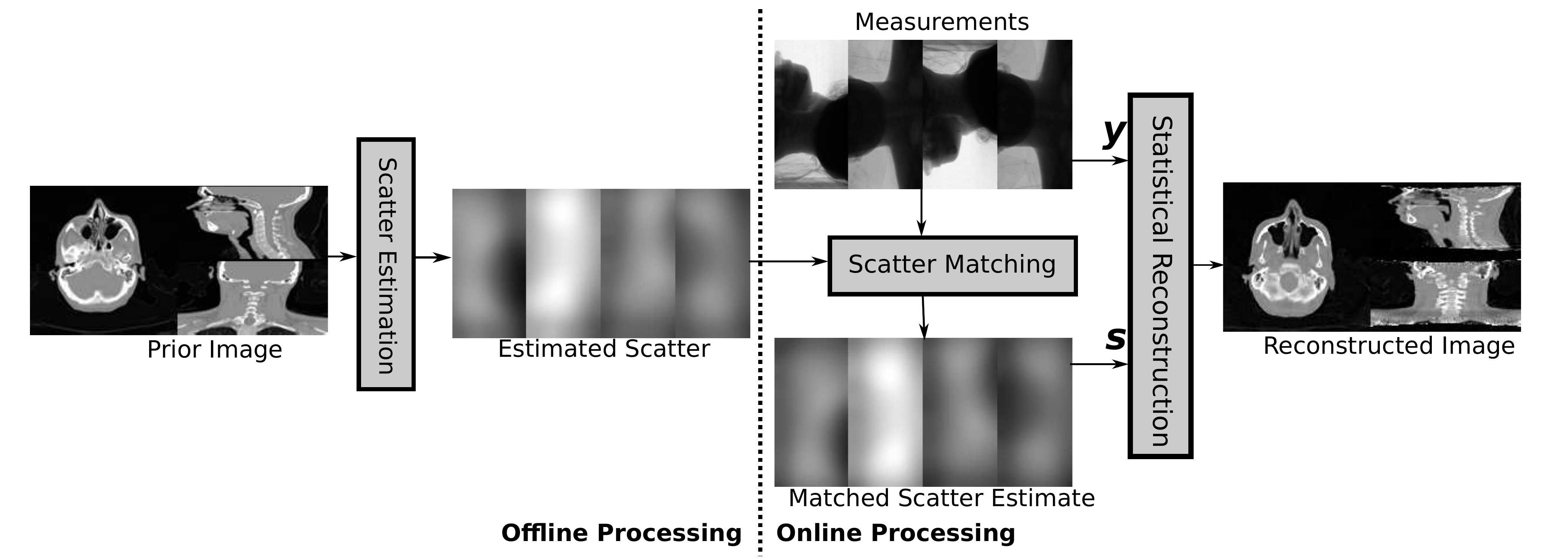}
	\caption{Flow diagram for off-line prior MC scatter estimation and statistical reconstruction.} 
	\label{fig:flow}
\end{figure}

We propose that the expectation of scatter may be calculated off-line to a high accuracy based upon a prior image, then matched to the measurements during replanning. Conceptually, this is very similar to the notion of SKS/ASKS \cite{Sun2010}, where the scatter point spread function of a scanner are measured through blocks of material, and combined with convolution. Instead here, the entire global scatter profile is estimated, and simply shifted to fit the current pose of the patient. Our framework for this off-line scheme is illustrated in Figure~\ref{fig:flow}.

For the match, we seek a transformation in the coordinates of detecting elements at each projection angle, for which we adopt the notation
\begin{equation} \label{equ:scat_match}
s_M\{i\} = \mathcal{I}\left(s\{i\};(u_M,v_M)_i\right) \mbox{ for }  i=1,\dots,P,
\end{equation}
where $P$ is the number of projections, $\mathcal{I}(\cdot)$ is the 2D linear interpolation of the image $s\{i\}$ corresponding to the $i^\mathrm{th}$ projection angle of scatter estimate, $s_M\{i\}$ is the matched estimate, and $(u_M,v_M)_i$ are the transformed 2D coordinates according to
\begin{equation} \label{equ:transform}
\left[\begin{array}{c}
u' \\
v' \\
1
\end{array}\right]
=
\begin{bmatrix}
w_{1,1} & w_{1,2} & w_{1,3} \\
w_{2,1} & w_{2,2} & w_{2,3} \\
0 & 0 & 1
\end{bmatrix}
\left[\begin{array}{c}
u \\
v \\
1
\end{array}\right],
\end{equation}
where $u$ and $v$ are the original vector of coordinates for the detector, and $w$ are parameters we wish to calculate through the matching process.

\begin{figure}
	\centering
	\includegraphics[width=0.5\linewidth]{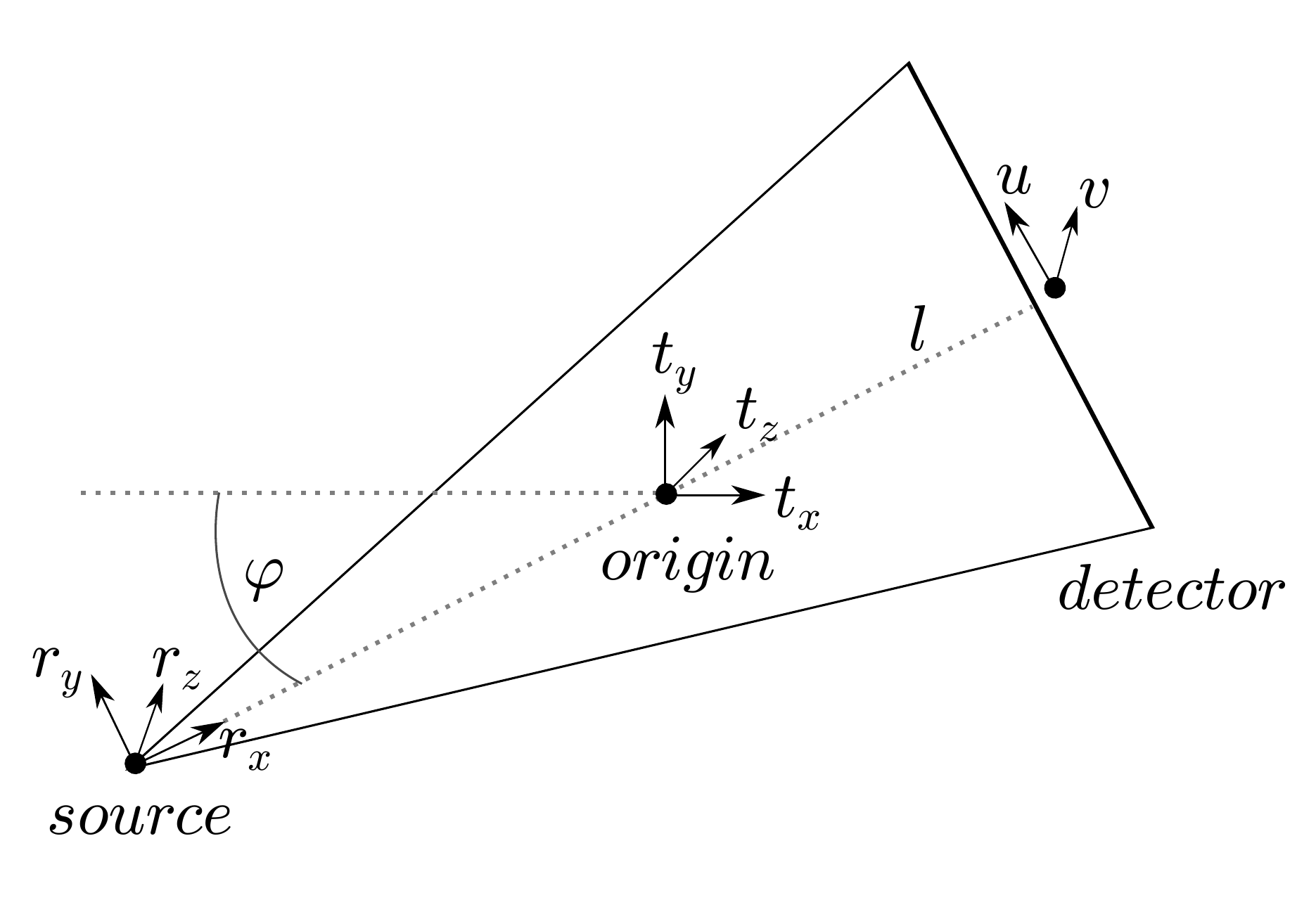}
	\caption{Geometry of scatter shifting.} 
	\label{fig:geometry}
\end{figure}

With reference to Figure~\ref{fig:geometry}, each scatter projection image with coordinates $u,v$ is updated based on a rigid translation of an object at the centre of rotation by $[t_x,t_y,t_z]^T$ onto a FDK reconstruction of the measurements. We define a new set of coordinates $[r_x,r_y,r_z]^T$, due to rotation of the source around by the $i^\mathrm{th}$ angle $\varphi_i$ by
\[
\left[\begin{array}{c}
r_x \\
r_y \\
r_z
\end{array}\right]
=
\begin{bmatrix}
\cos\varphi_i & -\sin\varphi_i & 0 \\
\sin\varphi_i & \cos\varphi_i & 0 \\
0 & 0 & 1
\end{bmatrix}
\left[\begin{array}{c}
t_x \\
t_y \\
t_z
\end{array}\right].
\]
We note that due to the projection, a translation along $r_x$ leads to a change in scaling, and translations in $r_y,r_z$ lead to shifting. If we define the distances $l_\mathrm{SO},l_\mathrm{SD}$ as the lengths from source to origin and the detector respectively, the transformation to adjust the projection is
\[
\left[\begin{array}{c}
u_M \\
v_M \\
1
\end{array}\right]
=
\begin{bmatrix}
\frac{l_\mathrm{SO}}{l_\mathrm{SO}+r_x} & 0 & \frac{l_\mathrm{SD}}{l_\mathrm{SO}+r_x}r_y \\
0 & \frac{l_\mathrm{SO}}{l_\mathrm{SO}+r_x} & \frac{l_\mathrm{SD}}{l_\mathrm{SO}+r_x}r_z \\
0 & 0 & 1
\end{bmatrix}
\left[\begin{array}{c}
u \\
v \\
1
\end{array}\right],
\]

\subsection{Reconstruction with Scatter Estimate} \label{sec:scat_est}
Instead using the approximate linearisation of the model in (\ref{equ:stat_model}), one could use it exactly. This is done for general additive Poisson noise in \cite{Erdogan1998}, and we will repeat it here explicitly for the incorporation of a scatter expectation in CBCT. In this case, reconstruction is taken as the maximum likelihood given (\ref{equ:stat_model}), with a regularisation function to impose desirable properties in the image, as in PWLS (\ref{equ:pwls}).

As in \cite{Erdogan1998}, we pursue finding the maximum likelihood, by minimising the negative log-likelihood of (\ref{equ:stat_model}), which is denoted
\begin{equation} \label{equ:nll}
\mathrm{NLL}(\boldsymbol{\mu};\boldsymbol{y},\boldsymbol{s}) =
\sum_{i=1}^{N}b_i\exp(-\left[\boldsymbol{\Phi}\boldsymbol{\mu}\right]_i)+s_i - y_i\log\left(b_i\exp(-\left[\boldsymbol{\Phi}\boldsymbol{\mu}\right]_i)+s_i\right).
\end{equation}
It should be noted that for $s_i>0$, (\ref{equ:nll}) is nonconvex, so it may not be minimised with the same ease of the PWLS. Nevertheless, it is continuously differentiable with respect to $\boldsymbol{\mu}$, and can therefore be treated with an appropriate first order method. We note that reconstruction is then solution of
\begin{equation} \label{equ:nll2}
\hat{\boldsymbol{\mu}} = \argmin_{\boldsymbol{\mu}\in\mathcal{C}} \mathrm{NLL}(\boldsymbol{\mu};\boldsymbol{y},\boldsymbol{s}) + \lambda R(\boldsymbol{\mu}),
\end{equation}
where $\mathcal{C}$ is a set of box constraints on $\boldsymbol{\mu}$ so that $0\leq\mu_i\leq \zeta \mbox{ for }  i=1,...,N$, where $\zeta$ is the maximum allowable attenuation coefficient.

Although some may consider the difference between our notions of `correction' with PWLS and `estimation' to be trivial, there is a compelling distinction. Whilst in the corrective case, one must carefully design the process to well approximate the model used, in estimation, the expectation of scatter can be used directly, and reconstruction may be considered as the direct inference from the raw measurements. How this translates into practical reconstruction accuracy will be studied in the experimental section.

\section{Experimentation}

\subsection{Data}
\begin{figure}[!ht]
	\centering
	\begin{subfigure}[b]{0.3\textwidth}
		\includegraphics[width=\textwidth]{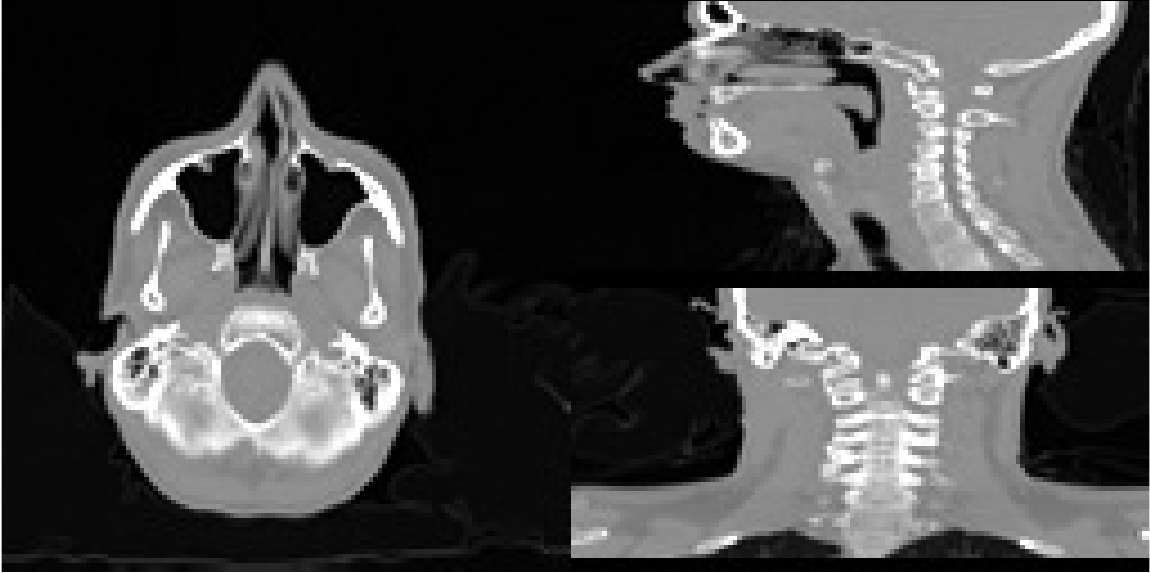}
		\caption{Oracle image}
		\label{subfig:im_oracle}
	\end{subfigure}
	\begin{subfigure}[b]{0.3\textwidth}
		\includegraphics[width=\textwidth]{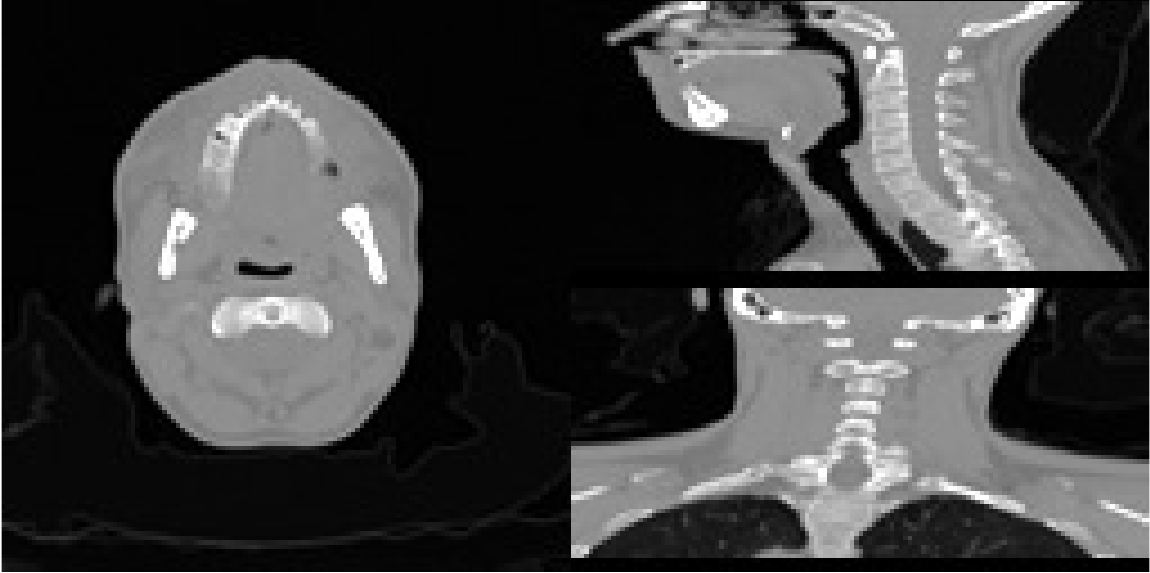}
		\caption{Planning image}
		\label{subfig:im_plan}
	\end{subfigure}
	\begin{subfigure}[b]{0.3\textwidth}
		\includegraphics[width=\textwidth]{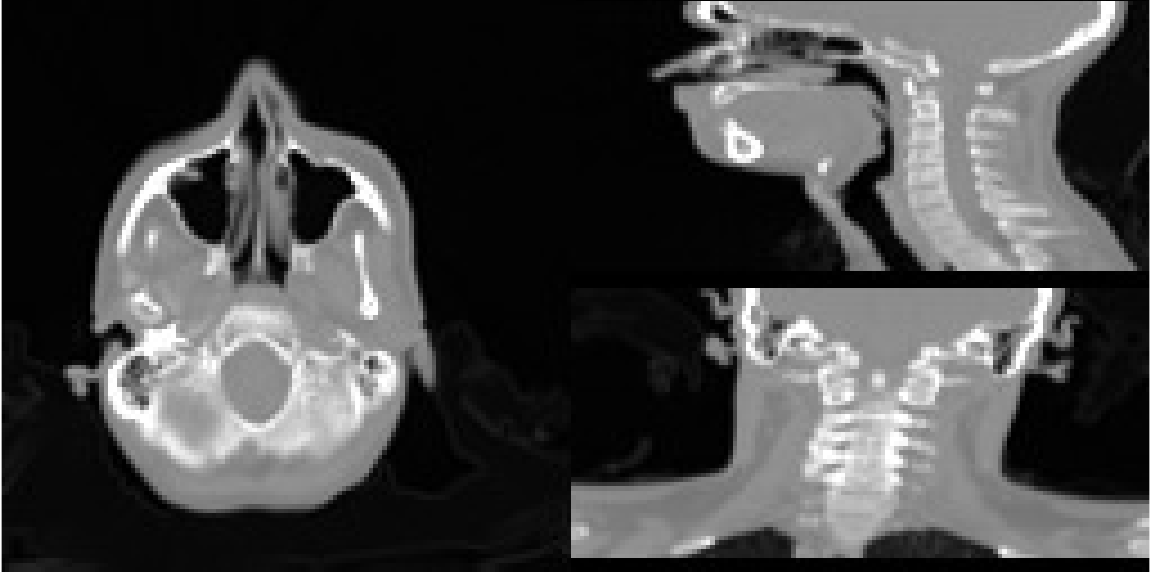}
		\caption{Registered plan}
		\label{subfig:im_reg}
	\end{subfigure}
	\caption{Experimental data used: (a) is the oracle follow-up CT image; (b) is an unregistered initial planning image; and (c) is the plan registered rigidly onto an FDK (with SKS correction) reconstruction of the raw data---shown is the high dose, but a separate registration was used in low dose.}
	\label{fig:data}
\end{figure}

The data set we are using is derived from repeat CT scans of a neck cancer patient from the Cancer Image Archive \cite{Clark2013}\cite{Fedorov2016}. With these, we will use the first CT scan as the planning image, then synthesise CBCT measurements on the follow up after 5 months---these are shown in Figures~\ref{subfig:im_oracle} and \ref{subfig:im_plan} respectively. A strong advantage of using this approach is that one has access to a ground truth, against which one can perform valid quantitative assessments.

To generate the CBCT data, we used the Monte-Carlo simulation tool Gate \cite{Jan2011} with a 60 keV monoenergetic source on the oracle image, where we did runs with $5\times10^{10}$ and $1\times10^{10}$ photons over 160 projection angles to represent two levels of dose.  

\subsection{Methods Under Test}
\subsubsection{Scatter Method Implementation}
\begin{figure}[!ht]
	\centering
	\begin{subfigure}[b]{0.3\textwidth}
		\includegraphics[width=\textwidth]{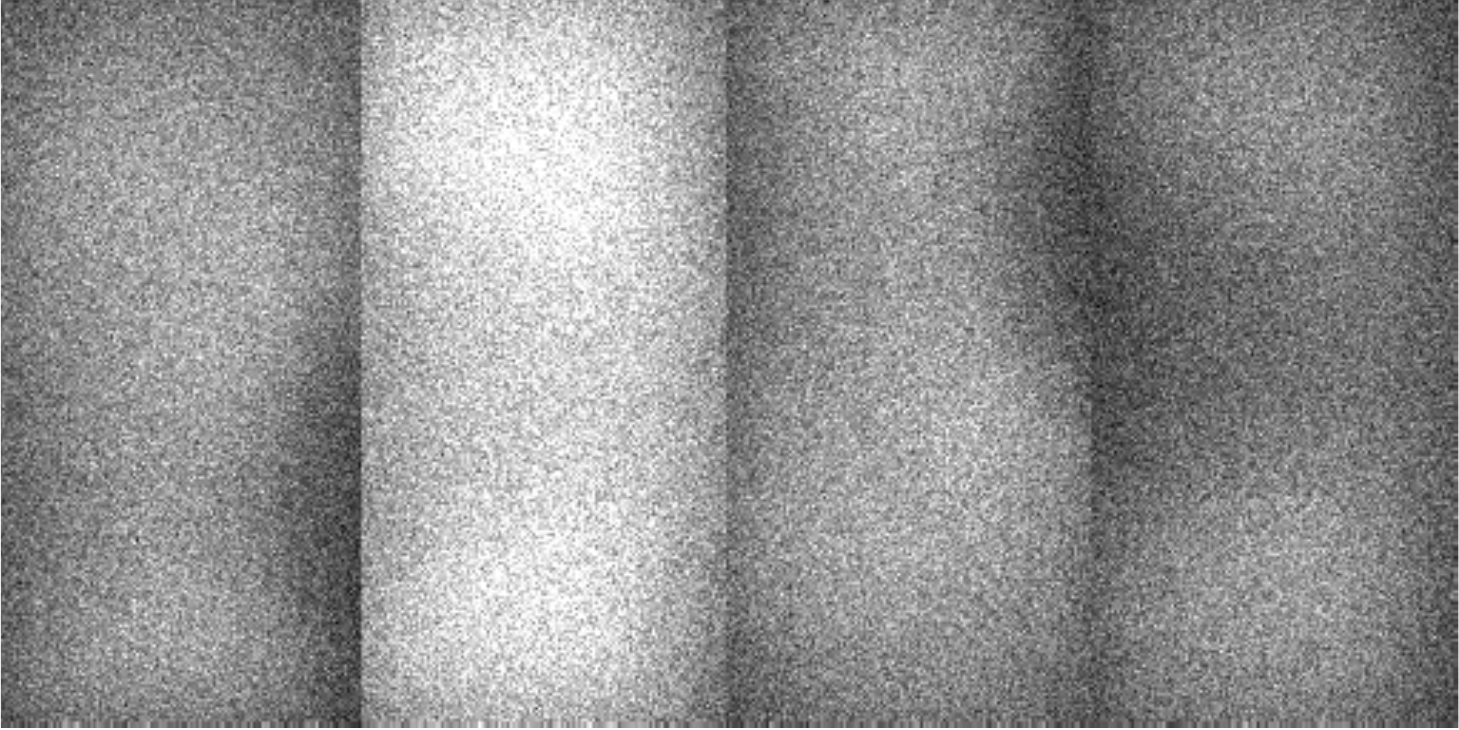}
		\caption{Oracle scatter}
		\label{subfig:scat_oracle}
	\end{subfigure}
	\begin{subfigure}[b]{0.3\textwidth}
		\includegraphics[width=\textwidth]{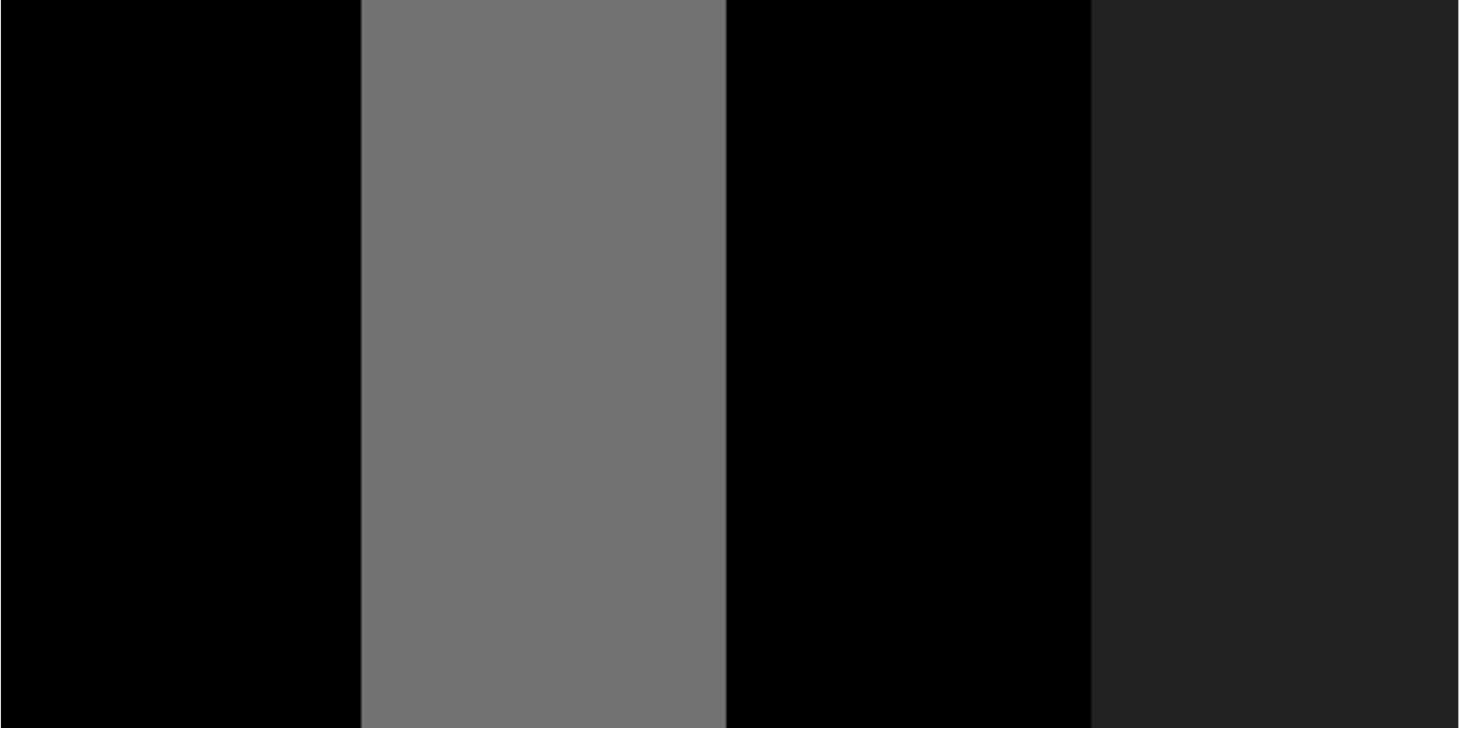}
		\caption{Uniform estimate}
		\label{subfig:scat_uni}
	\end{subfigure}
	\begin{subfigure}[b]{0.3\textwidth}
		\includegraphics[width=\textwidth]{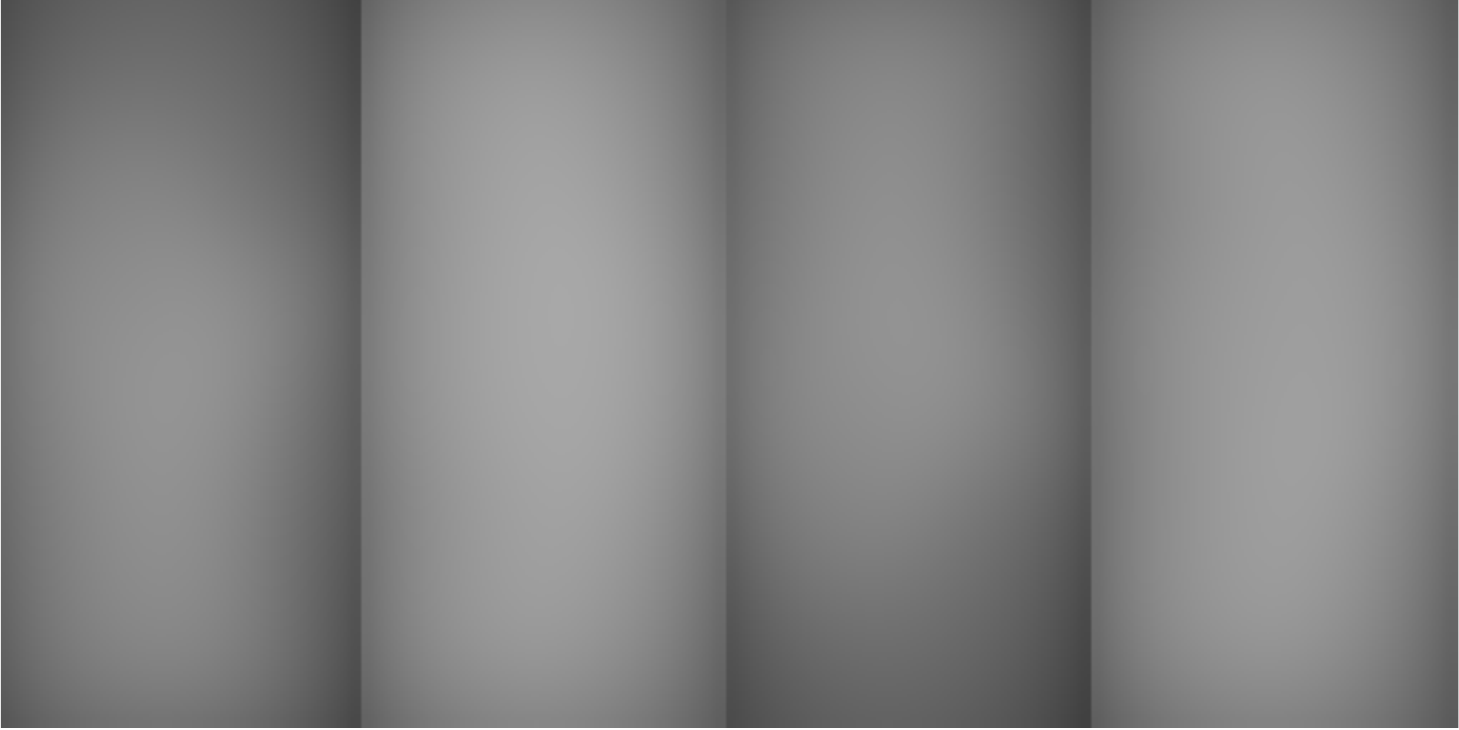}
		\caption{SKS estimate}
		\label{subfig:scat_sks}
	\end{subfigure}
	\vfil
	\begin{subfigure}[b]{0.3\textwidth}
		\includegraphics[width=\textwidth]{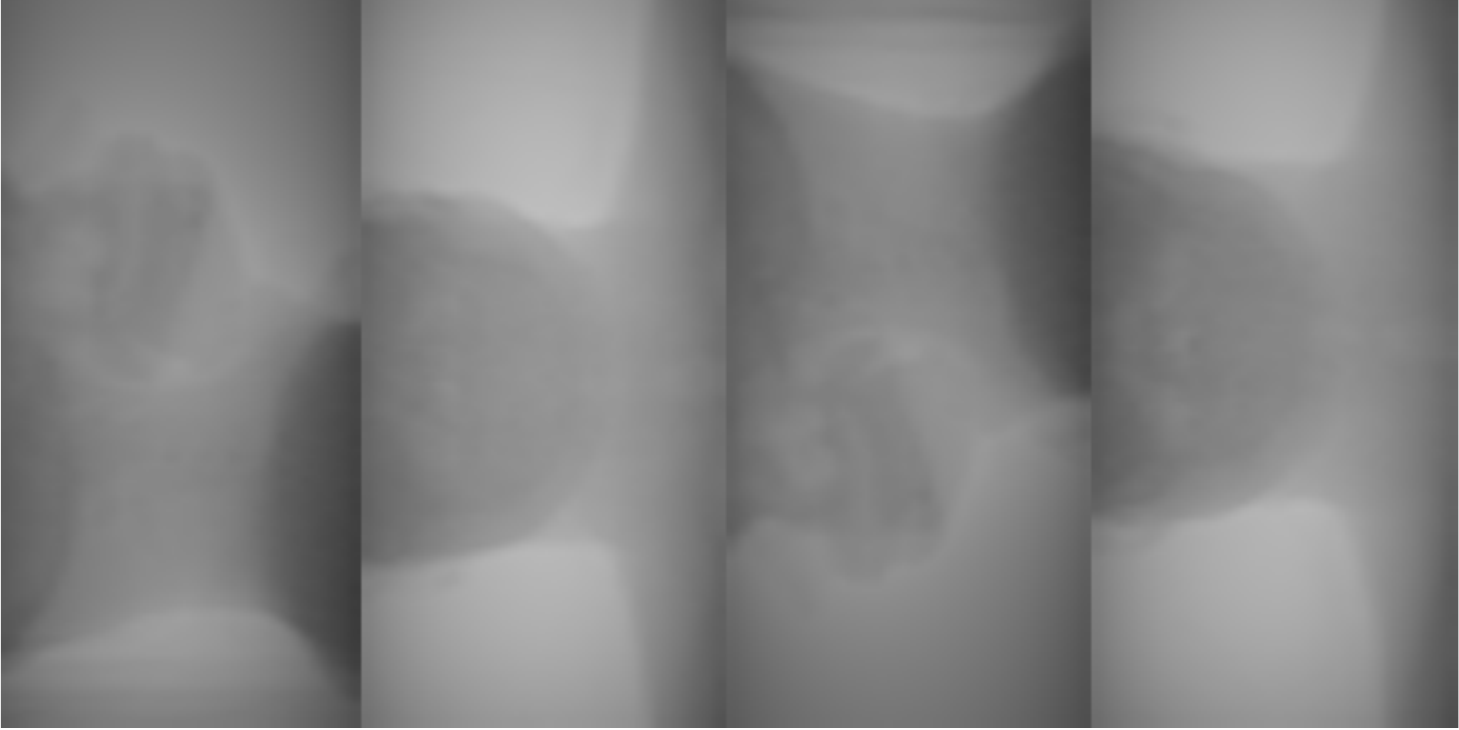}
		\caption{ASKS estimate}
		\label{subfig:scat_asks}
	\end{subfigure}
	\begin{subfigure}[b]{0.3\textwidth}
		\includegraphics[width=\textwidth]{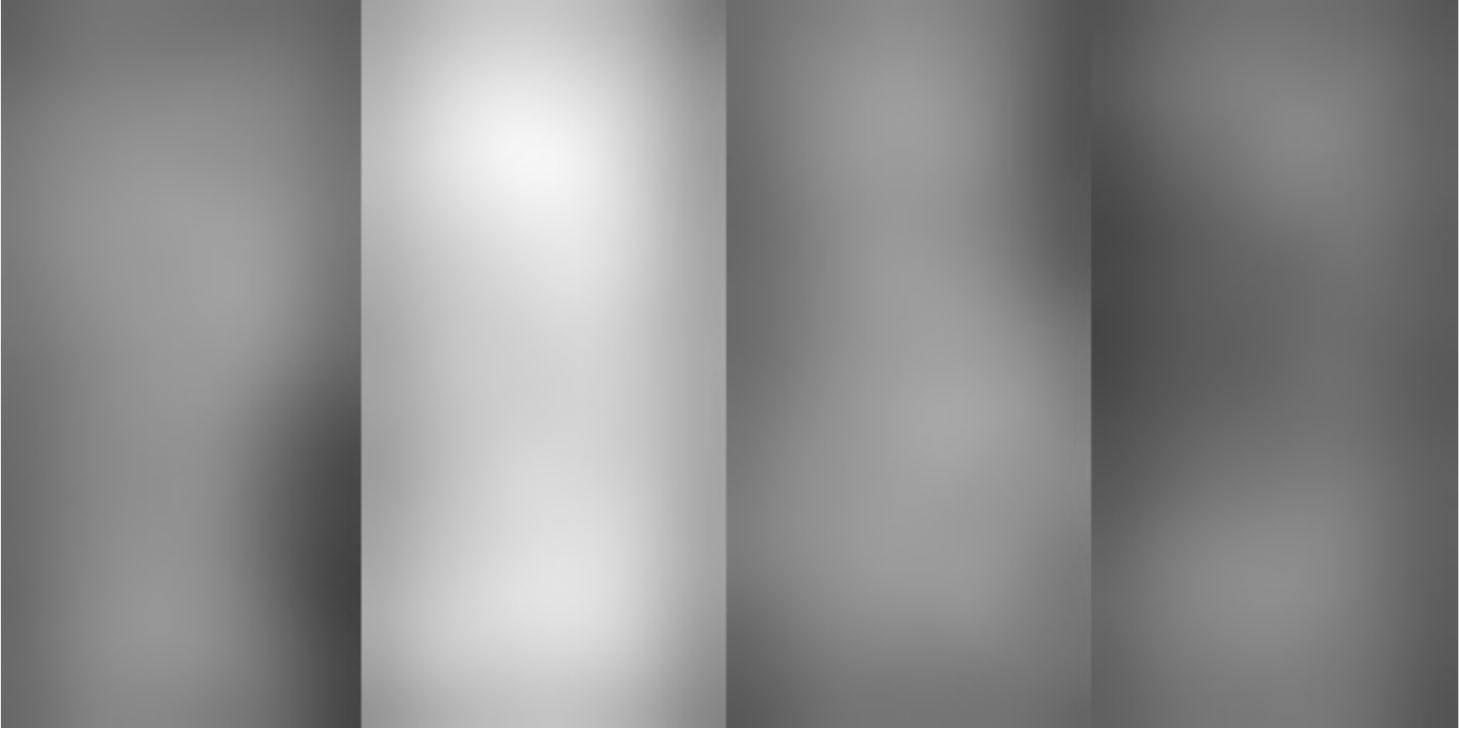}
		\caption{Online-prior-MC}
		\label{subfig:scat_reg}
	\end{subfigure}
	\begin{subfigure}[b]{0.3\textwidth}
		\includegraphics[width=\textwidth]{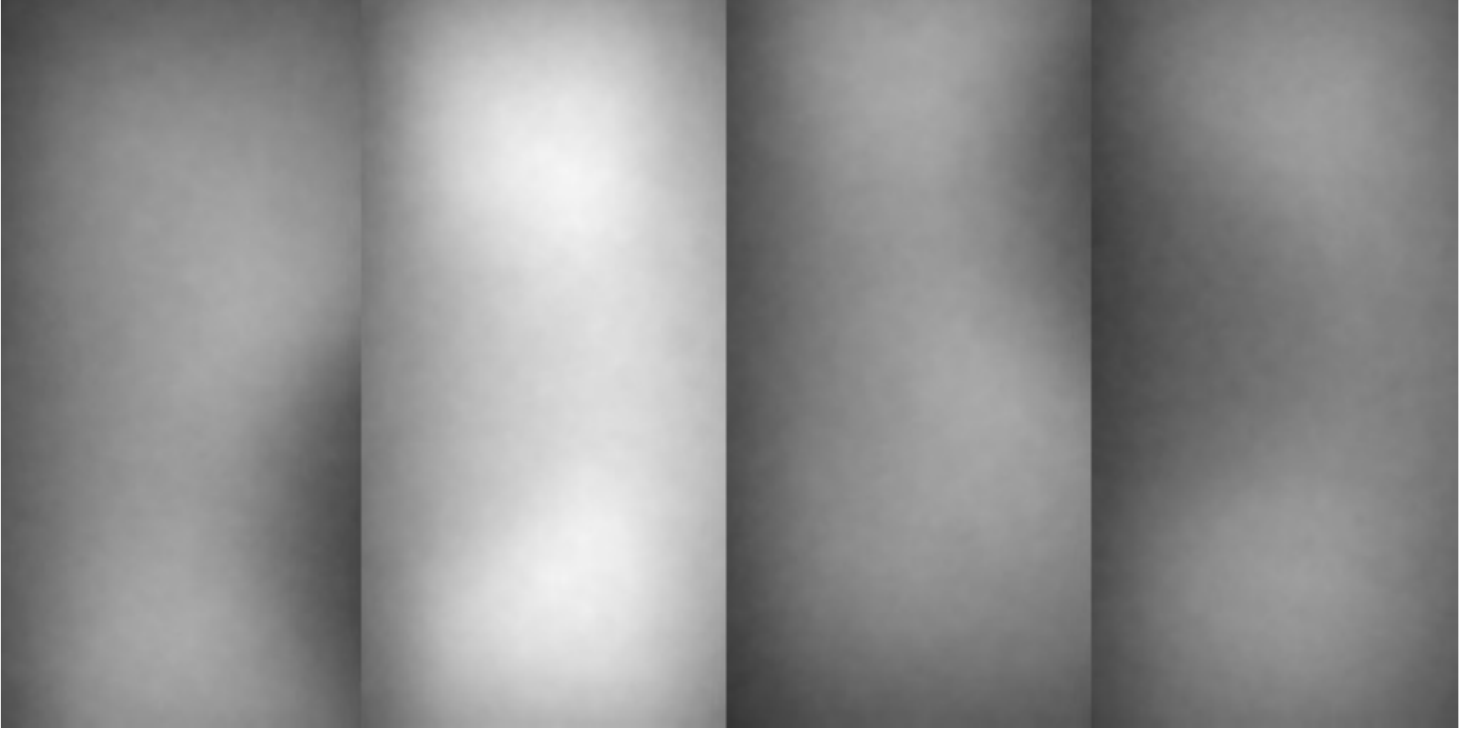}
		\caption{Offline-prior-MC}
		\label{subfig:scat_tc}
	\end{subfigure}

	\caption{Examples of low-dose scatter estimates shown with grey scale [10,70]: (a) is the oracle scatter from the measurement synthesis; (b) is a uniform estimate; (c) and (d) are SKS and fASKS estimates respectively; (e) and (f) are on-line and off-line planning MC estimates respectively.}
	\label{fig:scat}
\end{figure}

\begin{itemize}
	\item Oracle: this is using the true scatter signal from the measurement synthesis, to represent the ultimate conceivable estimate, and ground-truth for assessment.
	\item None: scatter signal is not estimated at all.
	\item Uniform: calculated using (\ref{equ:uniform}) \cite{Boellaard1997} with $\mathrm{SPR}=0.04$ and $t_\mathrm{air}=4000,800$ for moderate and low doses.
	\item SKS/fASKS: implemented with same parameters as \cite{Sun2010} for `full-fan' acquisition and 20 iterations each.
	\item Diff. filt.: using (\ref{equ:diff_filt}) based upon the registered plan in Figure~\ref{subfig:im_reg}.
	\item FDK-MC: a sub-sampled MC estimate based upon preliminary FDK with SKS.
	\item Online-prior-MC: the work-flow in \cite{Xu2015}, with sub-sampled MC applied to the registered planning image in Figure~\ref{subfig:im_reg}.
	\item Offline-prior-MC: using a detailed MC of the unregistered planning image in Figure~\ref{subfig:im_plan}, and matching as in Section~\ref{sec:match} from same registration parameters as other planning methods.
\end{itemize}
For illustrative purposes, select scatter estimates, along with the ground-truth are shown in Figure~\ref{fig:scat}.

\subsubsection{Reconstruction Implementation}
The reconstruction methods under test were FDK, PWLS according to (\ref{equ:pwls}), and inference (NLL) according to (\ref{equ:nll2}). All iterative methods were run for 200 iterations, which was deemed ample for convergence, and all $\lambda$ in the case of PWLS and NLL was set to $2\times10^5$, which was numerically tuned for good performance in both cases.

\subsection{Results} \label{sec:results}
Results are summarised in Tables~\ref{tab:low_results} and \ref{tab:high_results}, and selected reconstruction images are shown in Table~\ref{fig:rec_results}.

\newcommand{\addpic}[1]{\includegraphics[width=0.33\textwidth]{#1}}
\newcolumntype{C}{>{\centering\arraybackslash}m{0.35\textwidth}}
\begin{table*}[ht!]\sffamily
	\begin{tabular}{l*2{C}@{}}
		\toprule
		Experiment & FDK & NLL \\ 
		\midrule
		oracle & \addpic{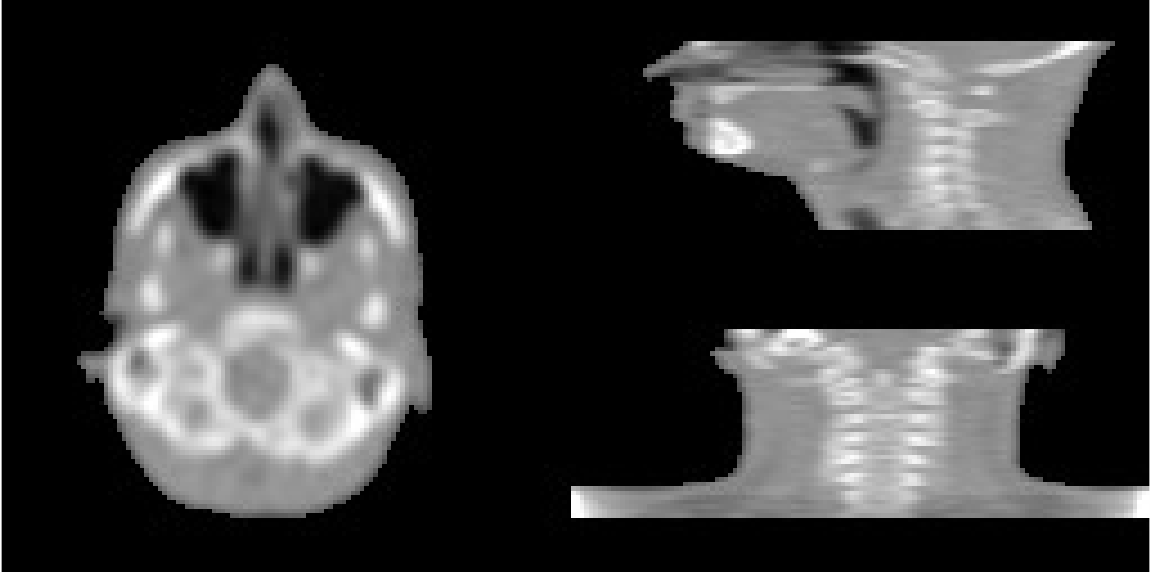} & \addpic{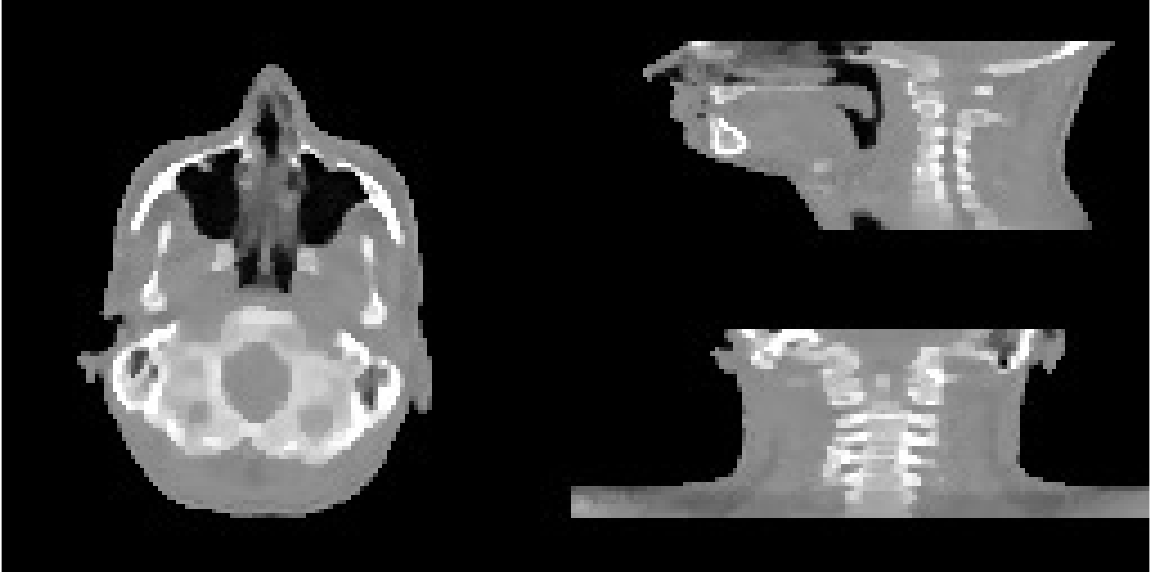}  \\ 
		none & \addpic{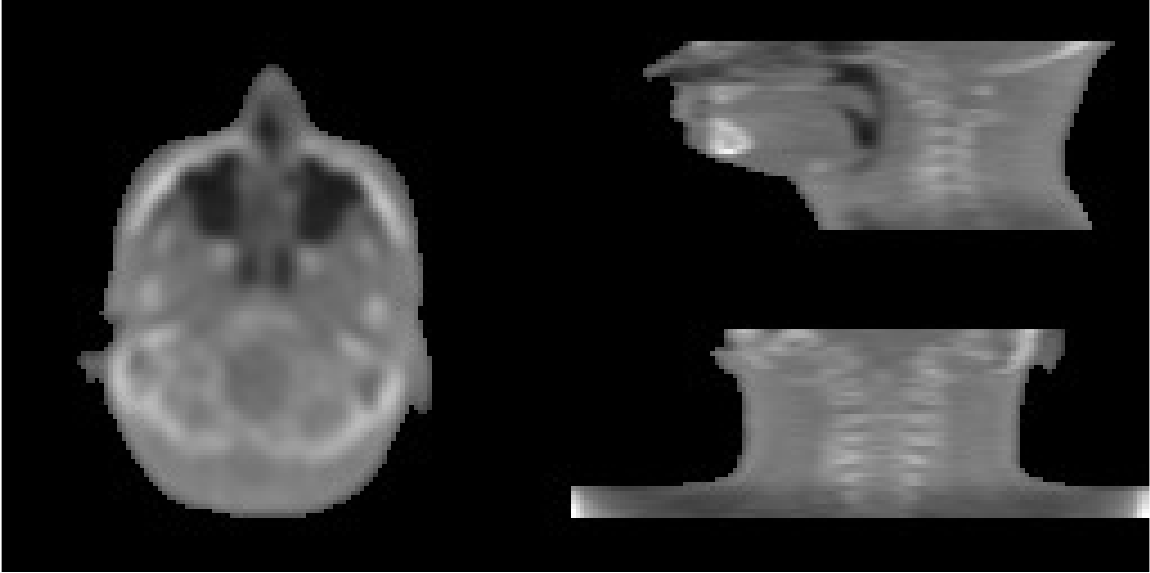} & \addpic{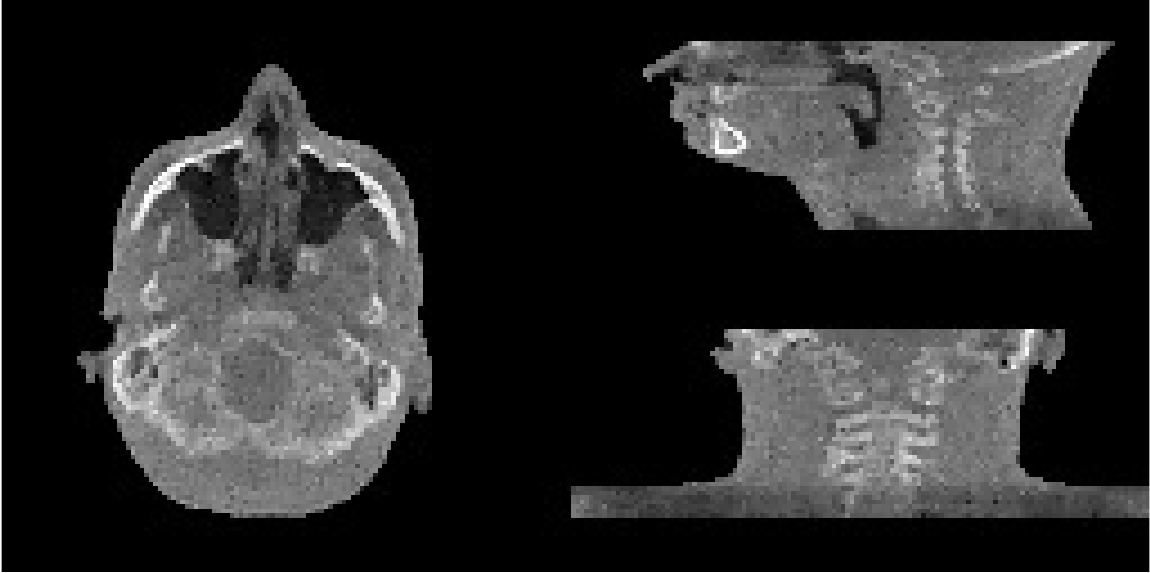}  \\ 
		SKS & \addpic{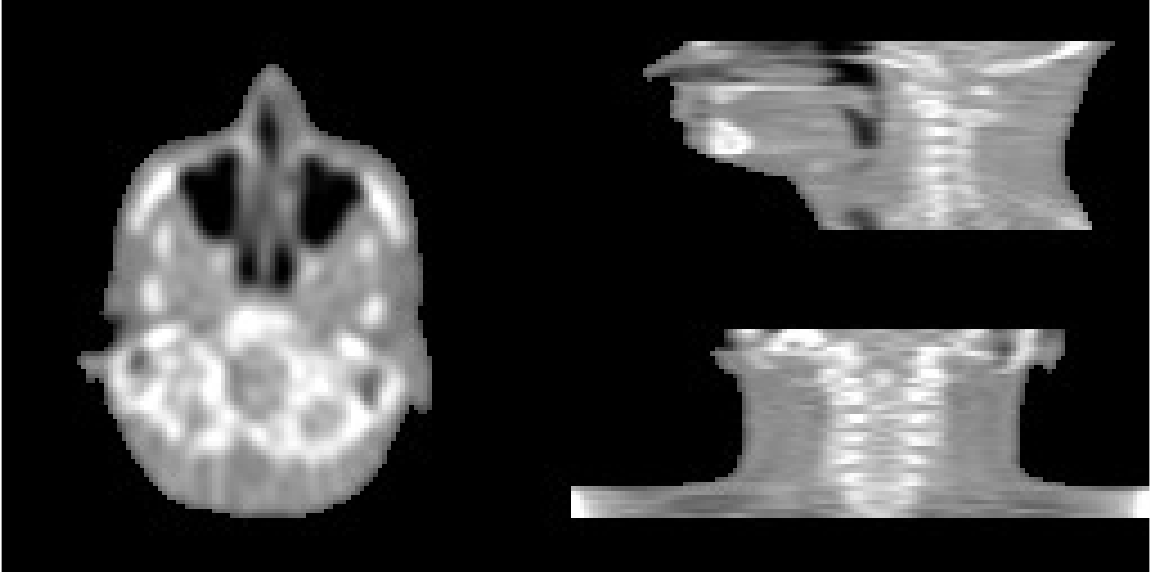} & \addpic{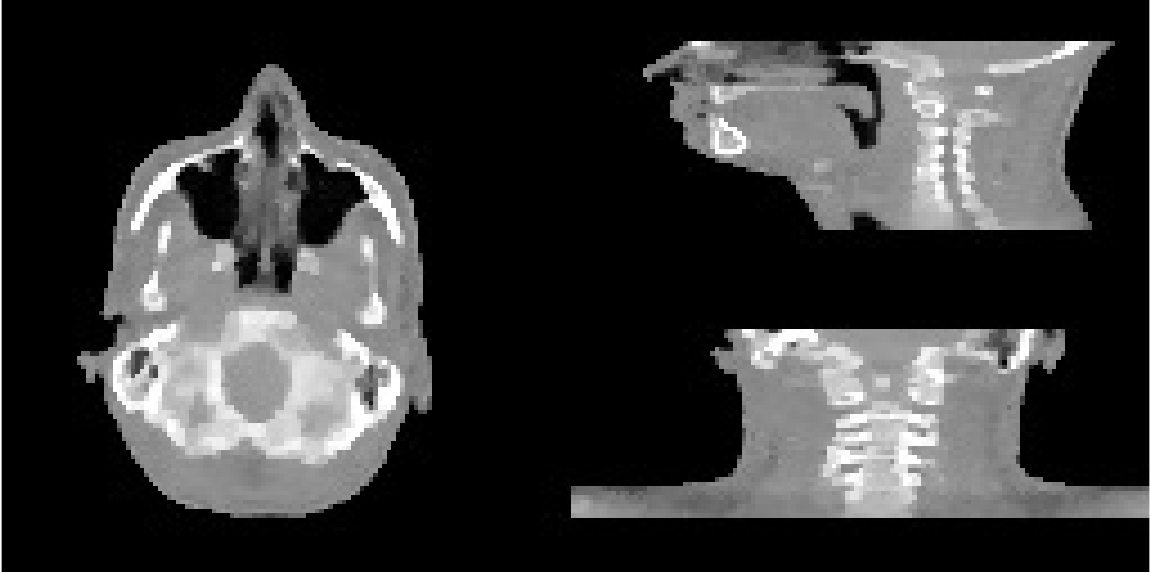}  \\ 
		offline-prior-MC & \addpic{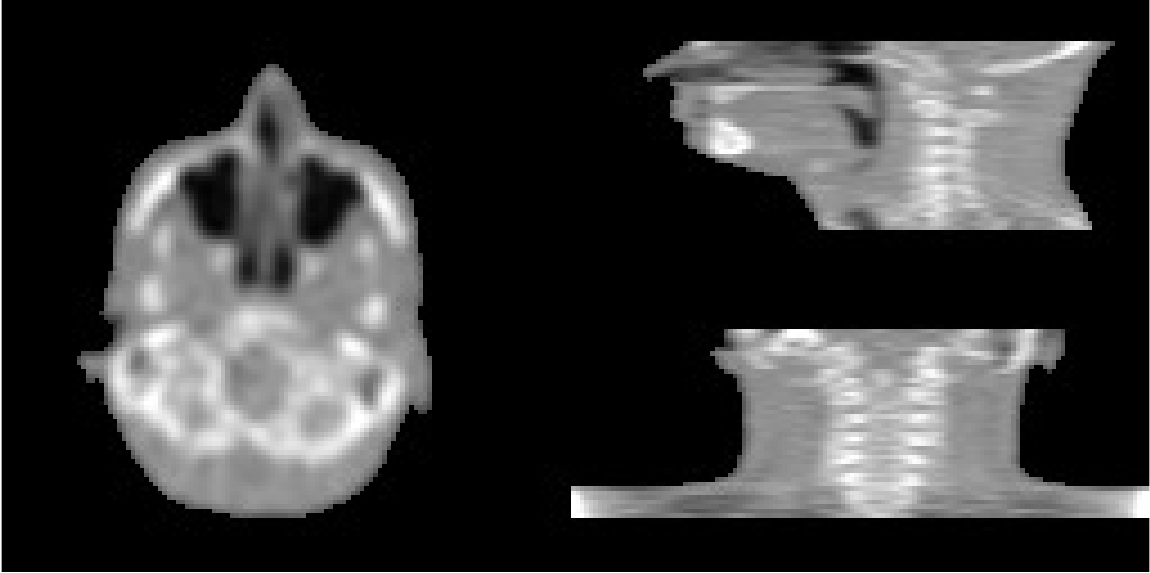} & \addpic{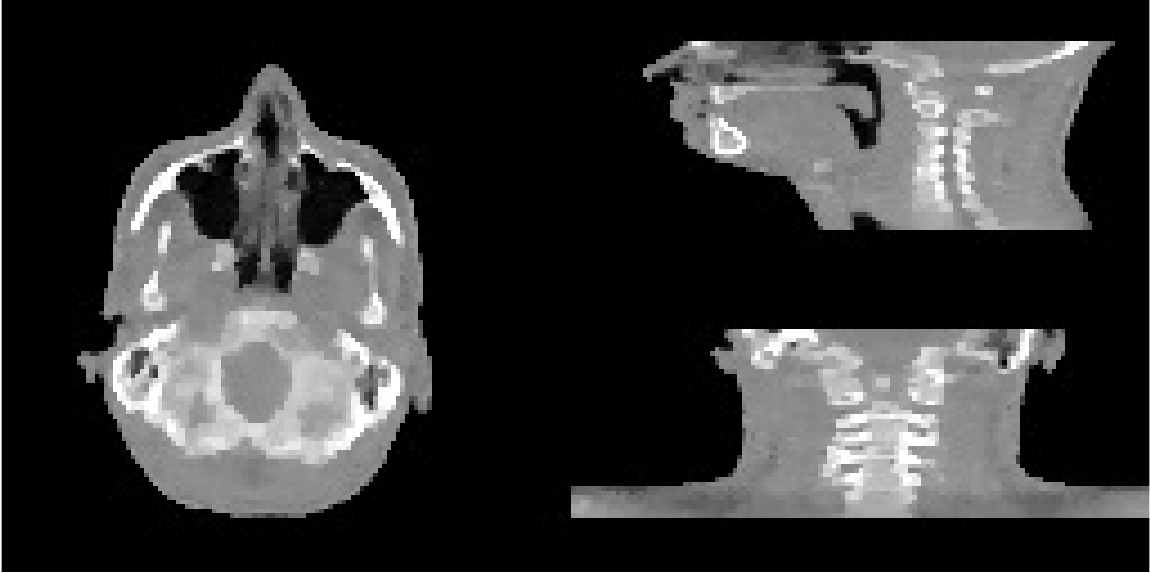}  \\ 
		\bottomrule 
	\end{tabular}
	\hspace{1cm}
	\caption{Experimental results from various estimation and reconstructions}
	\label{fig:rec_results}
\end{table*} 

\begin{table}[ht!]
	\caption{Quantitative results for high dose ($5\times10^5$ photons). All errors are given as root-mean-squared (RMS), and reconstruction errors are in Hounsfield Units.}
	\label{tab:high_results}
	\centering
	$
	\begin{array}{c|c|c|c|c}
	\hline
	\text{scheme} & \text{scatter error} & \text{FDK error} & \text{PWLS error} & \text{NLL error}\\
	\hline
	\text{oracle} & 0 & 51.8 & 19.8 & 19.3 \\
	\text{none} & 218 & 74.35 & 78.3 & 76.3 \\
	\hline
	\multicolumn{5}{c}{\text{measurement based online scatter calculation}}\\
	\hline
	\text{uniform} & 77.7 & 53.9 & 44.1 & 36.4 \\
	\text{SKS} & 38.7 & 60.3 & 24.2 & 21.7\\
	\text{ASKS} & 33.9 & 50.8 & 27.1 & 23.1\\
	\text{FDK-MC} & 102 & 56.0 & 46.0 & 40.1\\
	\hline
	\multicolumn{5}{c}{\text{prior based online scatter calculation}}\\
	\hline
	\text{filt. diff.} & 135 & 63.3 & 66.7 & 62.5 \\
	\text{online-prior-MC} & 18.5 & 54.1 & 22.5 & 21.2\\
	\hline
	\multicolumn{5}{c}{\text{prior based offline scatter calculation}}\\
	\hline
	\text{offline-prior-MC} & 24.6 & 58.1 & 22.7 & 21.8 \\
	\hline
	\end{array}
	$
\end{table}

\begin{table}[ht!]
	\caption{Quantitative results for low dose ($1\times10^5$ photons). All errors are given as root-mean-squared (RMS), and reconstruction errors are in Hounsfield Units}
	\label{tab:low_results}
	\centering
	$
	\begin{array}{c|c|c|c|c}
	\hline
	\text{scheme} & \text{scatter error} & \text{FDK error} & \text{PWLS error} & \text{NLL error}\\
	\hline
	\text{oracle} & 0 & 51.6 & 23.7 & 22.5 \\
	\text{none} & 43.9 & 74.1 & 78.5 & 77.4 \\
	\hline
	\multicolumn{5}{c}{\text{measurement based online scatter calculation}}\\
	\hline
	\text{uniform} & 27.3 & 71.9 & 65.0 & 60.6 \\
	\text{SKS} & 9.69 & 56.9 & 29.3 & 24.2\\
	\text{ASKS} & 8.95 & 51.5 & 34.1 & 26.0\\
	\text{FDK-MC} & 21.2 & 55.7 & 49.1 & 41.4\\
	\hline
	\multicolumn{5}{c}{\text{prior based online scatter calculation}}\\
	\hline
	\text{filt. diff.} & 27.8 & 63.5 & 44.7 & 43.2 \\
	\text{online-prior-MC} & 6.90 & 53.0 & 29.6 & 24.1\\
	\hline
	\multicolumn{5}{c}{\text{prior based offline scatter calculation}}\\
	\hline
	\text{offline-prior-MC} & 7.65 & 55.6 & 29.1 & 24.6 \\
	\hline
	\end{array}
	$
\end{table}

The first observation that can be made from the scatter accuracy in high and low dose cases, is that the prior-MC methods are the best. This is interestingly opposed to the FDK-MC, especially since this is based upon the SKS FDK, over which it has a significantly worse scatter estimate and only a slight decrease of analytic reconstruction accuracy. Another significant result is the very poor performance of the diff. filt. method, giving the worst estimates of scatter, which is likely due to large errors from mismatches between the registered plan. Perhaps this would decrease with a non-rigid registration as in \cite{Nui2010}, though this will inevitably be increasingly difficult and unstable in the lower dose settings.

In general, the relationship between relative errors in SKS/fASKS and the prior MC methods is enlightening. Although fASKS is the best performer in FDK, this does propagate through the rest. Apart from this however, the relative performance of these methods is very similar within the iterative results, all of which become rather close to the oracle scatter reconstruction in NLL. Of these, SKS may be the most appealing due to its fast computation and no reliance to planning registration.

One global trend in both the moderate and low dose results in Tables~\ref{tab:low_results} and \ref{tab:high_results} is that NLL is more accurate than PWLS on every count. This is unsurprising, since PWLS may be considered an approximation to NLL, but is motivating for pursuing inference methods rather than pre-corrected in the high scatter setting of CBCT. Furthermore, as the low reduces, the difference between these reconstruction models grows.

\section{Conclusions}
In this study, we have provided valid evidence for differences between various scatter estimation strategies, and how these may best be incorporated into reconstruction. The most conclusive message is that opting for the NLL is more accurate than pre-correcting for scatter and using the PWLS, and in a lower dose setting this difference becomes significant. In terms of scatter estimation, several of the methods aided rather accurate reconstruction: SKS, fASKS, and both on-line/off-line planning MC estimating. Which approach specifically to chose will be dependent on the application, and further evaluations may be valuable for other imaging scenarios and acquisitions.

\section{Acknowledgements}
The authors would like to thank the Maxwell Advanced Technology Fund, EPSRC DTP studentship funds and ERC project: C-SENSE (ERC-ADG-2015-694888) for supporting this work.
\clearpage
\bibliography{qcbct}

\begin{thebibliography}{10}

\bibitem{Button2010}
M~R Button and J~N Staffurth.
\newblock {Clinical Application of Image-guided Radiotherapy in Bladder and
  Prostate Cancer}.
\newblock {\em Clin. Oncol.}, 22(8):698--706, oct 2010.

\bibitem{Siewerdsen2001}
J~H Siewerdsen and D~A Jaffray.
\newblock {Cone-beam computed tomography with a flat-panel imager: Magnitude
  and effects of x-ray scatter}.
\newblock {\em Med. Phys.}, 28(2):220, 2001.

\bibitem{Poludniowski2009a}
G~Poludniowski, P~M Evans, V~N Hansen, and S~Webb.
\newblock {An efficient Monte Carlo-based algorithm for scatter correction in
  keV cone-beam CT}.
\newblock {\em Phys. Med. Biol.}, 54(12):3847--3864, 2009.

\bibitem{Love1987}
L~A Love and R~A Kruger.
\newblock {Scatter estimation for a digital radiographic system using
  convolution filtering}.
\newblock {\em Med. Phys.}, 14(2):178--185, mar 1987.

\bibitem{Boellaard1997}
R~Boellaard, M~van Herk, and B~J Mijnheer.
\newblock {A convolution model to convert transmission dose images to exit dose
  distributions}.
\newblock {\em Med. Phys.}, 24(2):189--199, feb 1997.

\bibitem{Sun2010}
M~Sun and J~M Star-Lack.
\newblock {Improved scatter correction using adaptive scatter kernel
  superposition}.
\newblock {\em Phys. Med. Biol.}, 55(22):6695--6720, nov 2010.

\bibitem{Nui2010}
T~Niu, M~Sun, J~Star-Lack, H~Gao, Q~Fan, and L~Zhu.
\newblock {Shading correction for on-board cone-beam CT in radiation therapy
  using planning MDCT images.}
\newblock {\em Med. Phys.}, 37(10):5395--406, oct 2010.

\bibitem{Marchant2008}
T~E Marchant, C~J Moore, C~G Rowbottom, R~I MacKay, and P~C Williams.
\newblock {Shading correction algorithm for improvement of cone-beam CT images
  in radiotherapy}.
\newblock {\em Phys. Med. Biol.}, 53(20):5719--5733, oct 2008.

\bibitem{Xu2015}
Y~Xu, T~Bai, H~Yan, L~Ouyang, A~Pompos, J~Wang, L~Zhou, S~B Jiang, and X~Jia.
\newblock {A practical cone-beam CT scatter correction method with optimized
  Monte Carlo simulations for image-guided radiation therapy}.
\newblock {\em Phys. Med. Biol.}, 60(9):3567--3587, may 2015.

\bibitem{Feldkamp1984}
L~A Feldkamp, L~C Davis, and J~W Kress.
\newblock {Practical cone-beam algorithm}.
\newblock {\em J. Opt. Soc. Am. A}, 1(6):612, 1984.

\bibitem{Park2015}
Y-K Park, G~C Sharp, J~Phillips, and B~A Winey.
\newblock {Proton dose calculation on scatter-corrected CBCT image: Feasibility
  study for adaptive proton therapy}.
\newblock {\em Med. Phys.}, 42(8):4449--4459, 2015.

\bibitem{Fessler2014}
J~A Fessler.
\newblock {Fundamentals of CT Reconstruction in 2D and 3D}.
\newblock In {\em Compr. Biomed. Phys.}, pages 263--295. Elsevier, 2014.

\bibitem{Elbakri2002}
I~A Elbakri and J~A Fessler.
\newblock {Statistical image reconstruction for polyenergetic X-ray computed
  tomography.}
\newblock {\em IEEE Trans. Med. Imaging}, 21(2):89--99, feb 2002.

\bibitem{Chang2014}
Z~Chang, R~Zhang, J-B Thibault, K~Sauer, and C~Bouman.
\newblock {Statistical x-ray computed tomography imaging from photon-starved
  measurements}.
\newblock {\em SPIE Comput. Imaging}, 9020:90200G, 2014.

\bibitem{Rit2014}
S~Rit, M~{Vila Oliva}, S~Brousmiche, R~Labarbe, D~Sarrut, and G~C Sharp.
\newblock {The Reconstruction Toolkit (RTK), an open-source cone-beam CT
  reconstruction toolkit based on the Insight Toolkit (ITK)}.
\newblock {\em J. Phys. Conf. Ser.}, 489:012079, mar 2014.

\bibitem{Erdogan1998}
H~Erdogan and J~A Fessler.
\newblock {Accelerated monotonic algorithms for transmission tomography}.
\newblock In {\em Proc. 1998 Int. Conf. Image Process. ICIP98 (Cat.
  No.98CB36269)}, volume~2, pages 680--684. IEEE Comput. Soc.

\bibitem{Clark2013}
K~Clark, B~Vendt, K~Smith, J~Freymann, J~Kirby, P~Koppel, S~Moore, S~Phillips,
  D~Maffitt, M~Pringle, L~Tarbox, and F~Prior.
\newblock {The Cancer Imaging Archive (TCIA): Maintaining and Operating a
  Public Information Repository}.
\newblock {\em J. Digit. Imaging}, 26(6):1045--1057, dec 2013.

\bibitem{Fedorov2016}
A~Fedorov, D~Clunie, E~Ulrich, C~Bauer, A~Wahle, B~Brown, M~Onken, J~Riesmeier,
  S~Pieper, R~Kikinis, J~Buatti, and R~R Beichel.
\newblock {DICOM for quantitative imaging biomarker development: a standards
  based approach to sharing clinical data and structured PET/CT analysis
  results in head and neck cancer research.}
\newblock {\em PeerJ}, 4:e2057, 2016.

\bibitem{Jan2011}
S~Jan, D~Benoit, E~Becheva, T~Carlier, F~Cassol, P~Descourt, T~Frisson,
  L~Grevillot, L~Guigues, L~Maigne, C~Morel, Y~Perrot, N~Rehfeld, D~Sarrut, D~R
  Schaart, S~Stute, U~Pietrzyk, D~Visvikis, N~Zahra, and I~Buvat.
\newblock {GATE V6: a major enhancement of the GATE simulation platform
  enabling modelling of CT and radiotherapy.}
\newblock {\em Phys. Med. Biol.}, 56(4):881--901, 2011.

\end{thebibliography}
\end{document}